\documentclass[a4paper,11pt]{article}
\usepackage[T1]{fontenc}

\usepackage[margin=1in]{geometry}
\usepackage{graphicx,libertine,afterpage,amsmath,amssymb,bm,textcomp,xcolor,soul,sfmath,natbib,authblk}

\usepackage{enumitem}

\definecolor{citecol}{rgb}{0,0.6,0}
\definecolor{linkcol}{HTML}{2171B5}
\usepackage[colorlinks=true,linkcolor=linkcol,citecolor=citecol,linktoc=all]{hyperref}

\usepackage[labelsep=period,labelfont=bf]{caption}
\usepackage[nameinlink]{cleveref}
\crefname{equation}{equation}{equations}
\crefrangelabelformat{equation}{#3#1#4-#5#2#6}
\creflabelformat{equation}{#2#1#3}

\title{\textcolor{red!50!black}{Asymptotic scaling properties of the posterior mean and variance in the Gaussian scale mixture model}}
\author{Rodrigo Echeveste, Guillaume Hennequin, M\'at\'e Lengyel} 
\date{\today}

\affil{\small Computational and Biological Learning Lab, Dept. of
Engineering, University of Cambridge, Cambridge, UK}

\providecommand{\argmax}{\mathop{\textrm{argmax}}}
\providecommand{\dd}{\mathrm{d}}

\providecommand{\fun}[2]{#1\!\left(#2\right)}
\providecommand{\funrm}[2]{\mathrm{#1}\!\left(#2\right)}

\providecommand{\transpose}[1]{#1^\mathsf{T}}
\providecommand{\E}[1]{\mathbb{E}\left[#1 \right]} 
\providecommand{\Cov}[1]{\mathbb{C}\left[#1 \right]} 

\providecommand{\PP}[1]{\mathcal{P}\!\left(#1\right)}

\providecommand{\QQ}[1]{\mathrm{Q}\!\left(#1\right)}

\providecommand{\given}{\vert}

\providecommand{\E}[1]{\mathrm{E}\!\left[#1\right]}

\providecommand{\cov}[1]{\mathrm{Cov}\!\left[#1\right]}

\providecommand{\normal}[1]{\fun{\mathcal{N}}{#1}}

\newcommand{\OO}{\boldsymbol{0}}

\newcommand{\II}{\mathbf{I}}
\newcommand{\ITL}{\tilde{\mathbf{I}}}

\newcommand{\xx}{\mathbf{x}}
\newcommand{\xxb}{\bar{\mathbf{x}}}
\newcommand{\xxbT}{\transpose{\bar{\mathbf{x}}}}
\newcommand{\xxT}{\transpose{\mathbf{x}}}
\newcommand{\yy}{\mathbf{y}}
\renewcommand{\dd}{\mathbf{d}}
\renewcommand{\AA}{\mathbf{A}}
\newcommand{\AAT}{\transpose{\mathbf{A}}}
\newcommand{\CC}{\mathbf{C}}
\newcommand{\DD}{\mathbf{D}}
\newcommand{\sx}{\sigma^2_\mathrm{x}}

\newcommand{\Nx}{N_\mathrm{x}}
\newcommand{\Ny}{N_\mathrm{y}}
\newcommand{\Nyns}{N_\mathrm{y_{\nosubs}}}

\newcommand{\mm}{\boldsymbol{\mu}}
\newcommand{\ZZ}{\boldsymbol{\Sigma}}

\renewcommand{\SS}{\mathbf{S}}

\newcommand{\subs}{\bullet}
\newcommand{\nosubs}{\circ}
\newcommand{\yys}{\yy_{\subs}}
\newcommand{\yyns}{\yy_{\nosubs}}
\newcommand{\AAs}{\AA_{\subs}}
\newcommand{\AAns}{\AA_{\nosubs}}
\newcommand{\AATns}{\AAT_{\nosubs}}
\newcommand{\CCs}{\CC_{\subs}}
\newcommand{\CCns}{\CC_{\nosubs}}
\newcommand{\CCsns}{\CC_{\nosubs\subs}}
\newcommand{\LL}{\mathbf{L}_{\nosubs}}
\newcommand{\LLT}{\transpose{\LL}}

\newcommand{\MM}{\mathbf{M}}
\newcommand{\TT}{\mathbf{T}}
\newcommand{\MMT}{\transpose{\MM}}

\newcommand{\VV}{\mathbf{V}}
\newcommand{\VVT}{\transpose{\VV}}

\newcommand{\VR}{\mathbf{V}_\rank}
\newcommand{\DR}{\mathbf{D}_\rank}
\newcommand{\VRT}{\transpose{\VR}}

\newcommand{\VRB}{\mathbf{V}_{\bar{\rank}}}
\newcommand{\VRBT}{\transpose{\VRB}}
\newcommand{\AAsp}{\bar{\AA}_{\subs}}
\newcommand{\RRsp}{\bar{\mathbf{R}}_{\nosubs}}

\newcommand{\lmms}{\mathbf{m}_{\subs}}
\newcommand{\lZZs}{\mathbf{S}_{\subs}}

\newcommand{\zs}{{z^\star}}
\newcommand{\zss}{\zs^{2}}
\newcommand{\zsc}{\zs^{3}}
\newcommand{\zsf}{\zs^{4}}
\newcommand{\zmap}{z_{\mathrm{MAP}}}

\newcommand{\mmz}{\mm^{z}}
\newcommand{\mms}{\mm_{\subs}}
\newcommand{\mmsz}{\mmz_{\subs}}
\newcommand{\mmsinf}{\mm^{\infty}_{\subs}}
\newcommand{\mmsinfT}{\mm^{\infty \mathsf{T}}_{\subs}}
\newcommand{\mmszero}{\mm^{0}_{\subs}}
\newcommand{\ZZs}{\ZZ_{\subs}}
\newcommand{\ZZz}{\ZZ^{z}}
\newcommand{\ZZsz}{\ZZz_{\subs}}

\newcommand{\ZZsinf}{\ZZ^{\infty}_{\subs}}
\newcommand{\ZZszero}{\ZZ^{0}_{\subs}}

\renewcommand{\QQ}{\mathbf{Q}_{\nosubs}}
\newcommand{\QQinf}{\QQ^{\infty}}
\newcommand{\QQb}{\overline{\mathbf{Q}}_{\nosubs}}

\newcommand{\RN}[1]{%
  \textup{\uppercase\expandafter{\romannumeral#1}}%
}

\newcommand{\ZZone}{\ZZ^{\RN{1}}}
\newcommand{\ZZtwo}{\ZZ^{\RN{2}}}

\newcommand{\rank}{\mathcal{R}}
\newcommand{\sz}[1]{\fun{\sigma^2_\mathrm{z}}{#1}}

\begin{document}

\maketitle

\tableofcontents

\parindent 0pt
\parskip 8pt
 
\section{Introduction}

The Gaussian scale mixture model (GSM) is a simple yet powerful probabilistic generative model of  natural image patches \citep{wainwright1999scale}. In line with the well-established idea that sensory processing is adapted to the statistics of the natural environment \citep{fiser10}, the GSM has also been considered a model of the early visual system, as a reasonable ``first-order'' approximation of the internal model that the primary visual cortex (V1) implements. According to this view, neural activities in V1 represent the posterior distribution under the GSM given a particular visual stimulus. Indeed, (approximate) inference under the GSM has successfully accounted for various nonlinearities in the mean (trial-average) responses of V1 neurons \citep{schwartz2001natural,coen2015flexible}, as well as the dependence of (across-trial) response variability with stimulus contrast found in V1 recordings \citep{orban2016neural}. However, previous work almost exclusively relied on numerical simulations to obtain these results. Thus, for a deeper insight into the realm of possible behaviours the GSM can (and cannot) exhibit and predict, here we present analytical derivations for the limiting behaviour of the mean and (co)variance of the GSM posterior at very low and very high contrast levels. These results should guide future work exploring neural circuit dynamics appropriate for implementing inference under the GSM.


\section{Recap: definition of the GSM}

\subsection{The generative model}

According to the GSM, an image patch $\xx \in \mathbb{R}^{\Nx}$ is constructed by linearly combining a (fixed) set of local features, $\AA \in \mathbb{R}^{\Nx\times\Ny}$, weighted by a set of (image-specific) coefficients, $\yy \in \mathbb{R}^{\Ny}$, and scaled by a single global (contrast) variable, $z\in \mathbb{R}$, plus additive white Gaussian noise:
\begin{align}
\xx \given \yy, z &\sim \normal{z \, \xxb, \sx\, \II} \text{, with } \xxb =  \AA \, \yy \label{eq:pred} \\
\intertext{where the feature coefficients are drawn from a multivariate Gaussian distribution}
\yy & \sim \normal{\OO, \CC} \label{eq:prior}
\end{align}
and the contrast, $z$, is drawn from a prior which we choose here to be a power-law:\footnote{In results to be presented elsewhere, we show how other popular choices for the prior, such as a gamma distribution or a truncated Gaussian, while leading to similar qualitative results within the range of interest for contrast, show divergent behaviour in the limit of very high contrast.}
\begin{align}
\PP{z} &= \frac{\left(n-1\right) \, b^{n-1}}{\left(z+b\right)^n} \label{eq:prior_z}
\end{align}
with $b>0$ and $n>1$.
It is the global contrast variable, $z$, which allows the model to produce higher-order statistical dependencies between local features, which are typically present in natural images \citep{schwartz2001natural}. 

\subsection{Posterior inference}

The posterior under the GSM for a given image, $\xx$, can be written as
\begin{align}
\PP{\yy \given \xx, z} &= \normal{\mm^z, \ZZ^z} \label{eq:posterior}\\
\text{with } \mm^z &= \frac{z}{\sx} \, \ZZ^z \, \AAT \, \xx \label{eq:postmean}\\
\text{and } \ZZ^z &=\left(\CC^{-1} + \frac{z^2}{\sx} \, \AAT \, \AA \right)^{-1} \label{eq:postvar}
\intertext{Where we have kept the superscript $\boldsymbol{\cdot}^z$ in the moments to explicitly denote their dependence on the inferred contrast $z$. In order to compute $\PP{\yy \given \xx}$, the moments of which we are ultimately interested  in (see above), we need to marginalise over $z$ using $\PP{z \given \xx}$:}
\PP{\yy \given \xx} &= \int \PP{\yy \given \xx, z} \, \PP{z \given \xx} \dd{z} \label{eq:margz}
\end{align}
While this posterior distribution is no longer a normal distribution, we can still compute its first two moments, given by:
\begin{align}
\mm &= \E{\mmz} \label{eq:postmean_full_inf}\\
\ZZ &= \E{\ZZz} + \Cov{\mm^z}  \label{eq:postvar_full_inf}
\end{align}
where $\E{\boldsymbol{\cdot}}$ and $\Cov{\cdot}$ denote expectation and covariance taken over $\PP{z \given \xx}$.

\section{Results}

We seek to understand how the mean and (co)variance of the posterior over feature coefficients, $\PP{\yy\given\xx}$, scale with contrast. In particular, we are interested in their asymptotic behaviour in the low- and high-contrast limits.

\subsection{Approximations to the inference of \texorpdfstring{$z$}{z}}\label{sec:inferz}

In the following, we distinguish between 
\begin{description}[align=left,labelwidth=3em,leftmargin=4em,itemindent=0em]
\item[$\zs$,] the true contrast of an image, such that we assume that the image, $\xx$, can be rewritten as a `base image', $\xxb$, scaled by this true contrast:\footnote{\label{fn:zinconsist}Note that this is slightly inconsistent with the generative model  (\Cref{eq:pred}), which assumes that observation noise gets added to the image after having been scaled by contrast.}
\begin{align}
\xx&=\zs\,\xxb \label{eq:scaleim}
\end{align}
\item[$z$,] the inferred contrast when that image is presented, i.e.\ the variable that needs to be inferred (and eventually marginalised out, see \Cref{eq:margz}) under the GSM;
\item[$\zmap$,] the maximum \emph{a posteriori} (MAP) estimate of the contrast, i.e.\ the setting of $z$ that maximises its posterior for a given image: $\zmap=\argmax_{z} \PP{z\given\xx}$.
\end{description}

We study two levels of approximation. First, we assume $\PP{z \given \xx} \simeq \fun{\delta}{z-\zmap}$, and therefore $\PP{\yy \given \xx} \simeq \PP{\yy \given \xx, z=\zmap}$ (see \Cref{eq:margz}) which makes $\PP{\yy \given \xx}$ a multivariate Gaussian (\Cref{eq:posterior}). Second, we also consider a further approximation, by assuming that $\zmap = \zs$, and thus $\PP{\yy \given \xx} \simeq \PP{\yy \given \xx, z=\zs}$, as in the limit of an infinitely large image patch the contrast should be near-perfectly inferred. However, these approximations prove to be too crude for computing the (co)variance of $\PP{\yy \given \xx}$ in the high contrast case, as they ignore the variance contributed by variability in $\mm^z$, the \emph{mean} of $\PP{\yy \given \xx, z}$, due to the non-zero variance of $\PP{z \given \xx}$ (\Cref{eq:postvar_full_inf}, second term). Thus, at high contrasts, we include this additional contribution by assuming that $\mm^z$ depends linearly on $z$ over the relevant range of $z$ around $\zmap$ (or $\zs$). (Note that the same linearity assumption means that no such correction is necessary for computing the mean of $\PP{\yy \given \xx}$.) Although, in principle, the same effect would also need to be considered at low contrasts, there the variance of $\PP{z \given \xx}$ is near zero, and so we will ignore it.

In summary, our strategy for analysing how the mean and (co)variance of $\PP{\yy \given \xx}$ scale with $\zs$ will proceed in two steps. We compute these quantities first as functions of $\zmap$ and then, via a mapping from $\zs$ to $\zmap$, as functions of $\zs$ (either computing the $\zs$-to-$\zmap$ mapping numerically, or, taking the second approximation, simply assuming an identity mapping). For brevity, we present below  (\Crefrange{sec:lowdim}{sec:lowhighcontrast}) the analytical results for the second, cruder approximation only, $\PP{\yy \given \xx} \simeq \PP{\yy \given \xx, z=\zs}$, and refer the reader to the Methods (\Cref{sec:methods}) for the analytical form of the first, milder approximation, $\PP{\yy \given \xx} \simeq \PP{\yy \given \xx, z=\zmap}$. We close this section by showing numerical results for both approximations (\Cref{sec:numerics}).

\subsection{Low-dimensional posterior}\label{sec:lowdim}

%
%
%
As we are only interested in low-order moments of the posterior (mean and [co]variance), we express the posterior for a subset of the latent variables, which we call $\yys$, with the rest of the latent variables denoted by $\yyns$ (such that $\yy=\left\{\yys,\yyns\right\}$). We denote the corresponding columns of $\AA$ by $\AAs$ and $\AAns$, and the corresponding blocks of $\CC$ by $\CCs=\cov{\yys}$, $\CCns=\cov{\yyns}$, and $\CCsns=\cov{\yyns,\yys}$. 

Thus, this low-dimensional posterior is (see \Cref{sec:lowpost} for a detailed derivation):
\begin{align}
\yys \given \xx, z &\sim \normal{\mmsz, \ZZsz} \label{eq:posteriors}\\
\text{where } \mmsz
&= \frac{z}{\sx} \, \ZZsz \, \transpose{\AAsp} \, \left( \frac{z^2}{\sx} \, \RRsp + \II \right)^{-1} \, \xx \label{eq:postmeans}\\
\text{and } \ZZsz 
&= \left[\CCs^{-1} + \frac{z^2}{\sx} \, \transpose{\AAsp} \left( \frac{z^2}{\sx} \, \RRsp + \II \right)^{-1} \, \AAsp \right]^{-1} \label{eq:postvars}\\
\text{with } \nonumber \\
\AAsp &= \AAs + \AAns \, \CCsns \, \CCs^{-1} \\
\RRsp &= \AAns \, \left(\CCns - \CCsns \, \CCs^{-1} \, \transpose{\CCsns} \right) \, \AATns 
\end{align}
(Note that \Crefrange{eq:posteriors}{eq:postvars} give back, as they should, \Crefrange{eq:posterior}{eq:postvar} in the special case when all of $\yy$ is included in $\yys$, and so $\AAns=\OO$ and thus $\RRsp=\OO$.)

\subsection{Low and high contrast limit scaling for the mean and the variance}\label{sec:lowhighcontrast}

\newcommand{\LC}{\mathrm{LC}}
\newcommand{\HC}{\mathrm{HC}}

In what follows we present scaling laws for the mean and the variance of the GSM, as a function of the (true) contrast variable $\zs$, both in the low contrast ($\LC$) limit ($\zs \rightarrow 0$) and in the high contrast ($\HC$) limit ($\zs \rightarrow \infty$). Up to second order in $\zs$, these take the general form (see \Cref{sec:methods}, Methods, for the derivations):
\begin{align}
\intertext{Low contrast:}
\fun{\mms^{\LC}}{\zs} &\simeq  \mmszero + \frac{\zss}{\sx} \, \MM^{\LC} \, \xxb  \label{eq:postmeans_LC}\\
\fun{\ZZs^{\LC}}{\zs} &\simeq \ZZszero - \frac{\zss}{\sx} \, \VV^{\LC} \label{eq:postvars_LC}\\
\intertext{High contrast:}
\fun{\mms^{\HC}}{\zs} &\simeq \mmsinf - \frac{\sx}{\zss} \, \MM^{\HC} \, \xxb  \label{eq:postmeans_HC}\\
\fun{\ZZs^{\HC}}{\zs} &\simeq \ZZsinf + \frac{\sx}{\zss} \, \VV^{\HC} \label{eq:postvars_HC}
\end{align}
Where $\mmszero$, $\mmsinf$, $\ZZszero$, $\ZZsinf$, $\MM^{\LC/\HC}$, $\VV^{\LC/\HC}$ are constant vectors and matrices, independent of the contrast level $\zs$, such that $\mmszero$, $\ZZszero$, $\mmsinf$, $\ZZsinf$ correspond to the asymptotic values of the mean and (co)variance at zero / infinite contrast, while $\MM^{\LC/\HC}$ and $\VV^{\LC/\HC}$ determine the speed of convergence towards these asymptotes. 
\Crefrange{eq:postmeans_LC}{eq:postvars_HC} reveal that in the low contrast regime, the magnitude of $\mms$ and $\ZZs$ grow quadratically. We also see that in the limit of infinite contrast, both the mean and variance decay towards their respective asymptotic values as $1/\zss$.

\textbf{In the low contrast limit,} we find (see \Cref{sec:methods} for details):
\begin{align}
\mmszero &= \bf{0}\\
\MM^{\LC} &=  \CCs \, \transpose{\AAsp} \\
\ZZszero &= \CCs \\
\VV^{\LC} &=  \CCs \, \transpose{\AAsp} \, \AAsp \, \CCs
\end{align}
Thus, the magnitude of the mean will grow quadratically from $0$ (which is the prior mean), while the variance (of a single unit) will decrease quadratically from the prior variance (because $\VV^{\LC}$ is positive definite, and so in the scalar case, it is positive).

\textbf{In the high contrast limit,} 
recall (\Cref{sec:inferz}) that there are two terms contributing to $\ZZs^{\HC}$: $\ZZsz$ at the fixed value of $z$ we are considering ($\zmap$ or $\zs$), and the the variance of $\mmsz$ due to posterior variability in $z$ around this fixed value. We will denote the corresponding terms in $\ZZsinf$ and $\VV^{\HC}$ by ${\ZZsinf}^{\RN{1}}$, ${\ZZsinf}^{\RN{2}}$ and  ${\VV^{\HC}}^{\RN{1}}$, ${\VV^{\HC}}^{\RN{2}}$, respectively, such that
\begin{align}
\ZZsinf &= {\ZZsinf}^{\RN{1}} + {\ZZsinf}^{\RN{2}} \\
\VV^{\HC} &= {\VV^{\HC}}^{\RN{1}} + {\VV^{\HC}}^{\RN{2}} \\
\intertext{where}
{\ZZsinf}^{\RN{2}} &= \frac{1}{2 \rank- n} \, \mmsinf \, \mmsinfT\\
{\VV^{\HC}}^{\RN{2}} &= \frac{3}{2 \rank- n} \, \left(\mmsinf \, \xxbT \, \transpose{{\MM^{\HC}}}  + \MM^{\HC} \, \xxb \, \mmsinfT \right)\\
\intertext{and}
\rank &= \funrm{rank}{\AA \CC \AAT}
\end{align}
(See \Cref{sec:var_mean} for a derivation.)

In order to derive the other coefficients, $\mmsinf$, $\MM^{\HC}$, ${\ZZsinf}^{\RN{1}}$, and ${\VV^{\HC}}^{\RN{1}}$, we note that the inverse of the matrix $\left( \frac{z^2}{\sx} \, \RRsp + \II \right)^{-1}$, present in both \Cref{eq:postmeans,eq:postvars}, imposes some limitations when $\RRsp$ is itself not invertible.
For an overcomplete model (formally, in which $\funrm{rank}{\AAns}=\Nx \leq \Nyns$), $\RRsp$ will be invertible. However, for an undercomplete model (which we will here restrict to the case $\funrm{rank}{\AAns}= \Nyns \leq \Nx$), $\RRsp$ will be low rank and thus non-invertible, in which case we will make use of its Cholesky decomposition:
\begin{align}
\RRsp&= \AAns \, \LL \, \LLT \, \AATns \label{eq:Rchol}\\
\intertext{where, in turn, $\LL$ is the Cholesky factor of  $\CCns - \CCsns \, \CCs^{-1} \, \transpose{\CCsns}$:}
\LL\,\LLT &= \CCns - \CCsns \, \CCs^{-1} \, \transpose{\CCsns}
\end{align}
With these considerations, we obtain separate solutions for the over- and undercomplete cases (see \Cref{sec:methods} for details).\footnote{A third case (not studied here) is also possible, in which $\funrm{rank}{\AAns} < \funrm{min}{\Nyns, \Nx}$ (see \Cref{sec:rank}).}

\begin{align}
\text{Overcomplete system:} \nonumber\\
{\ZZsinf}^{\RN{1}} & = \left(\CCs^{-1} \transpose{\AAsp} \, \RRsp^{-1}  \, \AAsp\right)^{-1}\\
{\VV^{\HC}}^{\RN{1}} &= {\ZZsinf}^{\RN{1}} \, \transpose{\AAsp} \, \RRsp^{-1}  \, \RRsp^{-1} \, \AAsp \, {\ZZsinf}^{\RN{1}} \\
\mmsinf &= {\ZZsinf}^{\RN{1}} \, \transpose{\AAsp} \, \RRsp^{-1} \, \xxb\\
\MM^{\HC} &= {\ZZsinf}^{\RN{1}} \, \transpose{\AAsp} \, \RRsp^{-1} \, \RRsp^{-1} \left( \II - \AAsp \, {\ZZsinf}^{\RN{1}} \, \transpose{\AAsp} \, \RRsp^{-1} \right)\\
~\\
\text{Undercomplete system:} \nonumber\\
{\ZZsinf}^{\RN{1}} & = \mathbf{0} \\
{\VV^{\HC}}^{\RN{1}} &= \left( \transpose{\AAsp} \, \QQinf \, \AAsp \right)^{-1} \\
\mmsinf &= {\VV^{\HC}}^{\RN{1}} \transpose{\AAsp} \, \QQinf \, \xxb\\
\MM^{\HC} &= {\VV^{\HC}}^{\RN{1}} \, \left[ \left(\CCs^{-1} +  \transpose{\AAsp} \, \QQb \, \AAsp \right) {\VV^{\HC}}^{\RN{1}} \, \transpose{\AAsp} \, \QQinf  - \transpose{\AAsp} \, \QQb \right]\\
\text{with } \nonumber \\
\QQinf &= \II - \AAns \, \LL \, \left( \LLT \, \AATns \, \AAns \, \LL \right)^{-1} \, \LLT \, \AATns \\
\QQb &= \AAns \, \LL \, \left( \LLT \, \AATns \, \AAns \, \LL \right)^{-2}  \, \LLT \, \AATns  
\end{align}

Note that ${\ZZsinf}^{\RN{1}}$ only shrinks to zero in the undercomplete but not in the overcomplete case. This is because, in the overcomplete case, the input is only able to pin down the value of the latent variables to a (linear) subspace even if $z$ is fixed, so the full posterior tends towards a rank-deficient (zero-thickness) `pancake' which marginalises to a full-rank (finite-volume) `cloud' when projected down to a low-dimensional subspace, thus leaving some ever-lingering uncertainty within that subspace. In contrast, in the undercomplete case, at any fixed contrast level $z$, the input actually overconstrains the latents (bar the effect of observation noise), and so the posterior over features tends towards a Dirac delta function which remains a delta function even after projecting down to a low-dimensional subspace. In this case, the only contribution to the variance  comes from ${\ZZsinf}^{\RN{2}}$.

\subsection{Numerical validation}\label{sec:numerics}

\begin{figure}[t]
\centering
\includegraphics[width=0.7\textwidth]{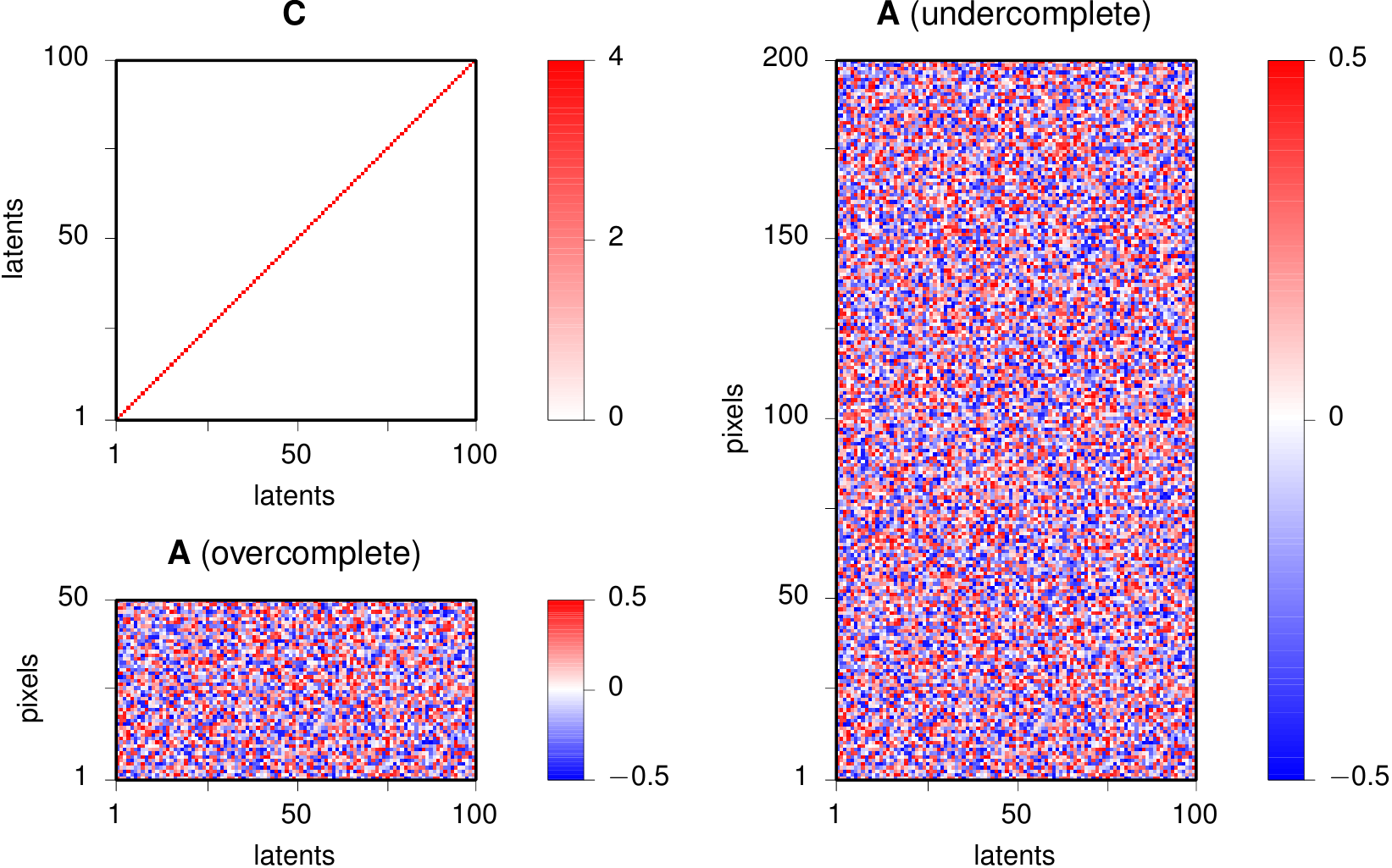}
\caption{\label{fig:matrices}
Identity prior covariance matrix $\CC$ (top left) together with random filter matrices $\AA$ used for numerical testing the low and high contrast approximations in the overcomplete (bottom left) and undercomplete cases (right). Note that the over- and undercomplete cases only differed in the number of observations (rows) but not of the latent variables (columns) so that the same prior could be applied.
}
\end{figure}

\begin{figure}[t]
\centering
\includegraphics[height=0.2\textheight]{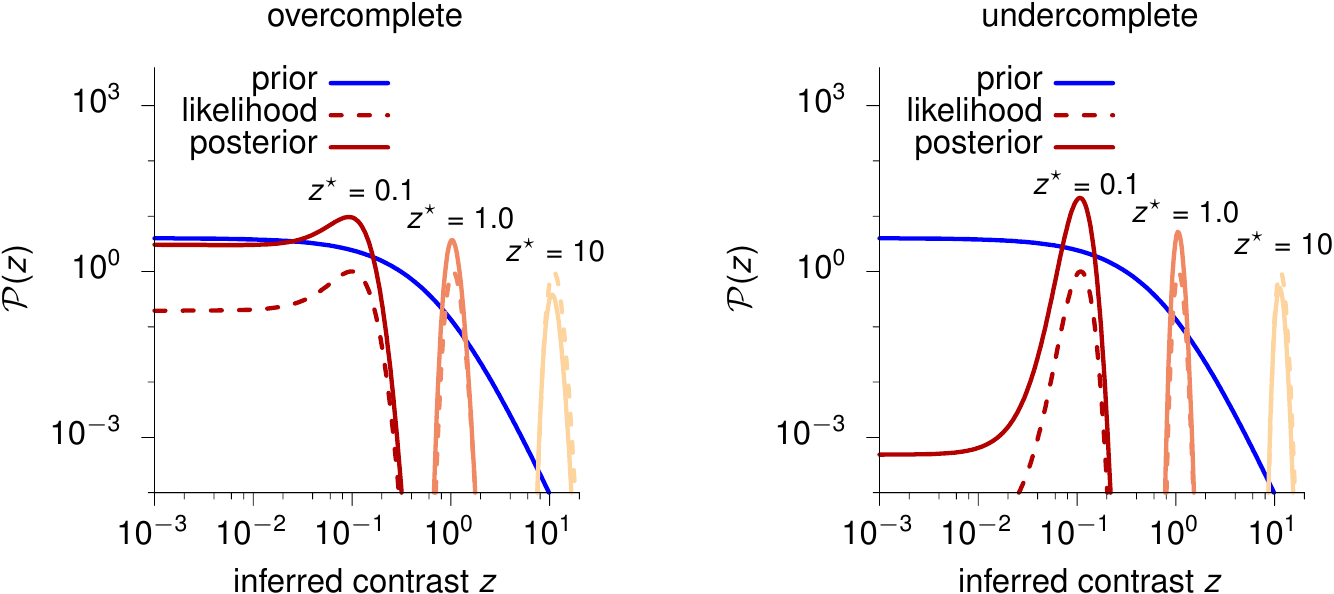}
\caption{\label{fig:prior_like_post}
Prior distribution over $z$ (blue), together with a few examples for the likelihood (shades of red, dashed) and posterior at different true contrast levels, $\zs$ (shades of red, solid), in the overcomplete (left) and undercomplete case (right).
}
\end{figure}

\begin{figure}[t]
\centering
\includegraphics[height=0.4\textheight]{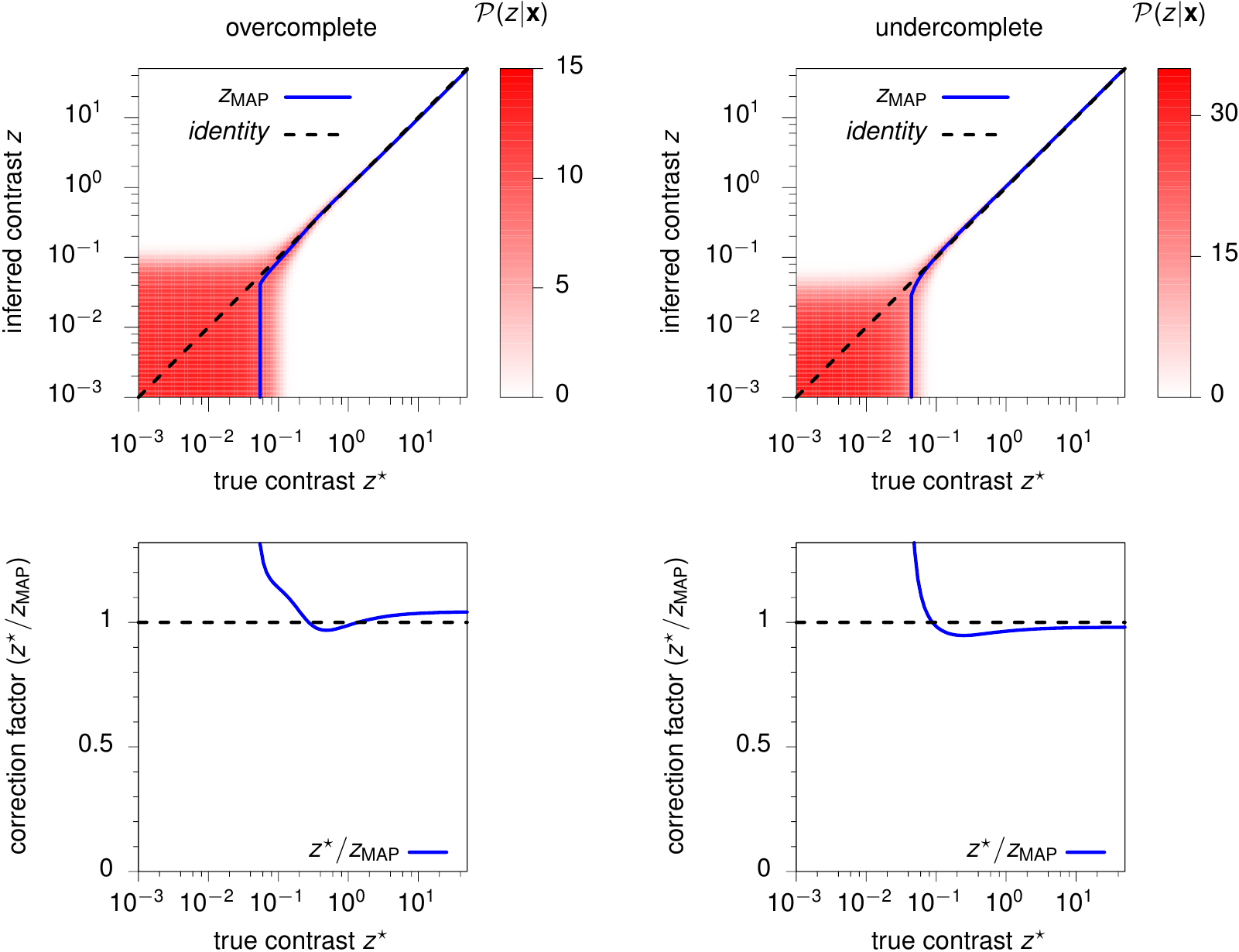}
\caption{\label{fig:z_post}
{\bfseries Top}:~posterior probability (color coded) of inferred contrast $z$ given an image, $\xx$, generated with true contrast $\zs$ (see text for details); the value of $\zmap$ as a function of $\zs$ is superimposed (blue lines).
{\bfseries Bottom}:~correction factors $\zs/\zmap$ as a function of  $\zs$, with dashed lines showing $\zmap = \zs$.
{\bfseries Left}:~overcomplete case. {\bfseries Right}:~undercomplete case.
}
\end{figure}

In order to test the quality of our approximations in \Crefrange{eq:postmeans_LC}{eq:postvars_HC}, we evaluated them together with the full expressions from \Crefrange{eq:posterior}{eq:margz} on a toy example. We chose an identity prior covariance $\CC$ (scaled by 4), a random filter matrix $\AA$ (each element sampled i.i.d.\ from a uniform between $-0.5$ and $0.5$), and fixed the observation noise level to $\sx=1$ (\Cref{fig:matrices}). 
We generated the input image, $\xx$, by sampling from the GSM (\Crefrange{eq:pred}{eq:prior}) with the parameters described above and $z=\zs$, which we varied systematically%
\footnote{This is slightly inconsistent with the simple scaling of input with $\zs$ that we assumed for the derivations, \Cref{eq:scaleim} (see also \Cref{fn:zinconsist}), but consistent with the generative model of the GSM and thus ensures e.g.\ that inferences about $z$ are well calibrated wrt.\ $\zs$.}.
(Specifically, to better isolate the effects of changing contrast, we used the same $\xxb$ and frozen observation noise in \Cref{eq:pred} for generating $\xx$ at all values of $\zs$).
%
To infer $z$, we used a power-law prior with $n = 4$ (such that both its mean and variance were finite) and $b = 0.75$. With these settings of the parameters, the posterior over $z$ was mostly dominated by the likelihood (\Cref{fig:prior_like_post}). In particular, it was unimodal and tight, with $\zmap$ following $\zs$ closely for all but the smallest true contrast levels (\Cref{fig:z_post}). The $z$-likelihood -- and thus the $z$-posterior -- was even tighter in the undercomplete case as it involved a higher number of observed variables (\Cref{fig:matrices}).

\begin{figure}[t]
\centering
\includegraphics[height=0.44\textheight]{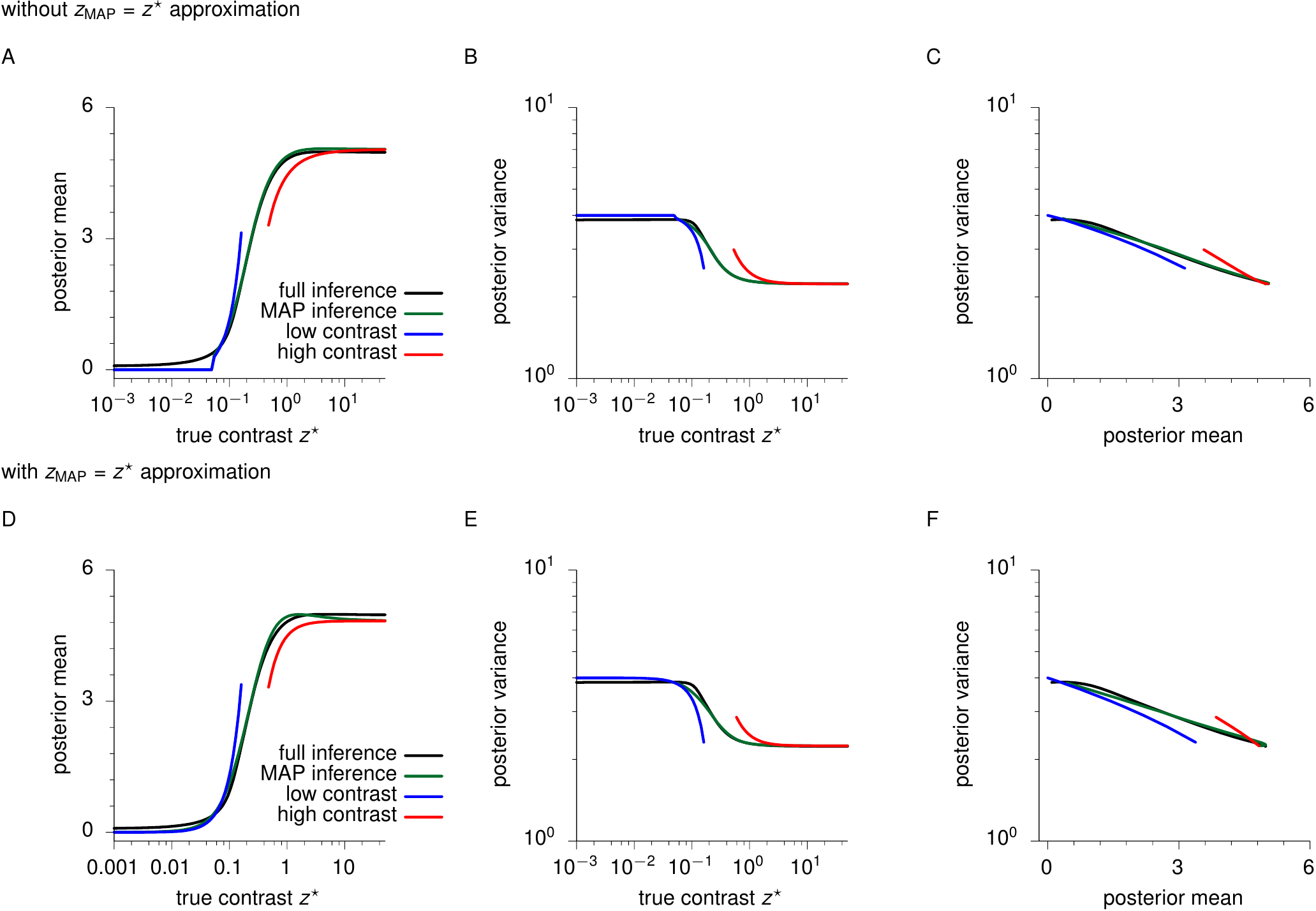}
\caption{\label{fig:overcomp}
{\bfseries Dependence of posterior mean and variance on contrast in the overcomplete case}. \textbf{A}-\textbf{B}, \textbf{D}-\textbf{E}: posterior mean (A, D) and variance (B, E), as functions of the true contrast level $\zs$. \textbf{C}, \textbf{F}: Posterior variance against posterior mean as $\zs$ is varied. All panels, black: results of `full' inference (obtained by numerically marginalising out the full $z$-posterior, \Crefrange{eq:posterior}{eq:margz}); green: MAP inference (by conditioning on $z=\zmap$ in \Crefrange{eq:postmean}{eq:postvar}); blue and red: low- and high-contrast approximations (\Crefrange{eq:postmeans_LC}{eq:postvars_HC}), respectively. Top panels (A, B, C): approximations using true $\zmap$; bottom panels (D, E, F): approximations substituting $\zmap$ with $\zs$. 
}
\end{figure}

\begin{figure}[t]
\centering
\includegraphics[height=0.44\textheight]{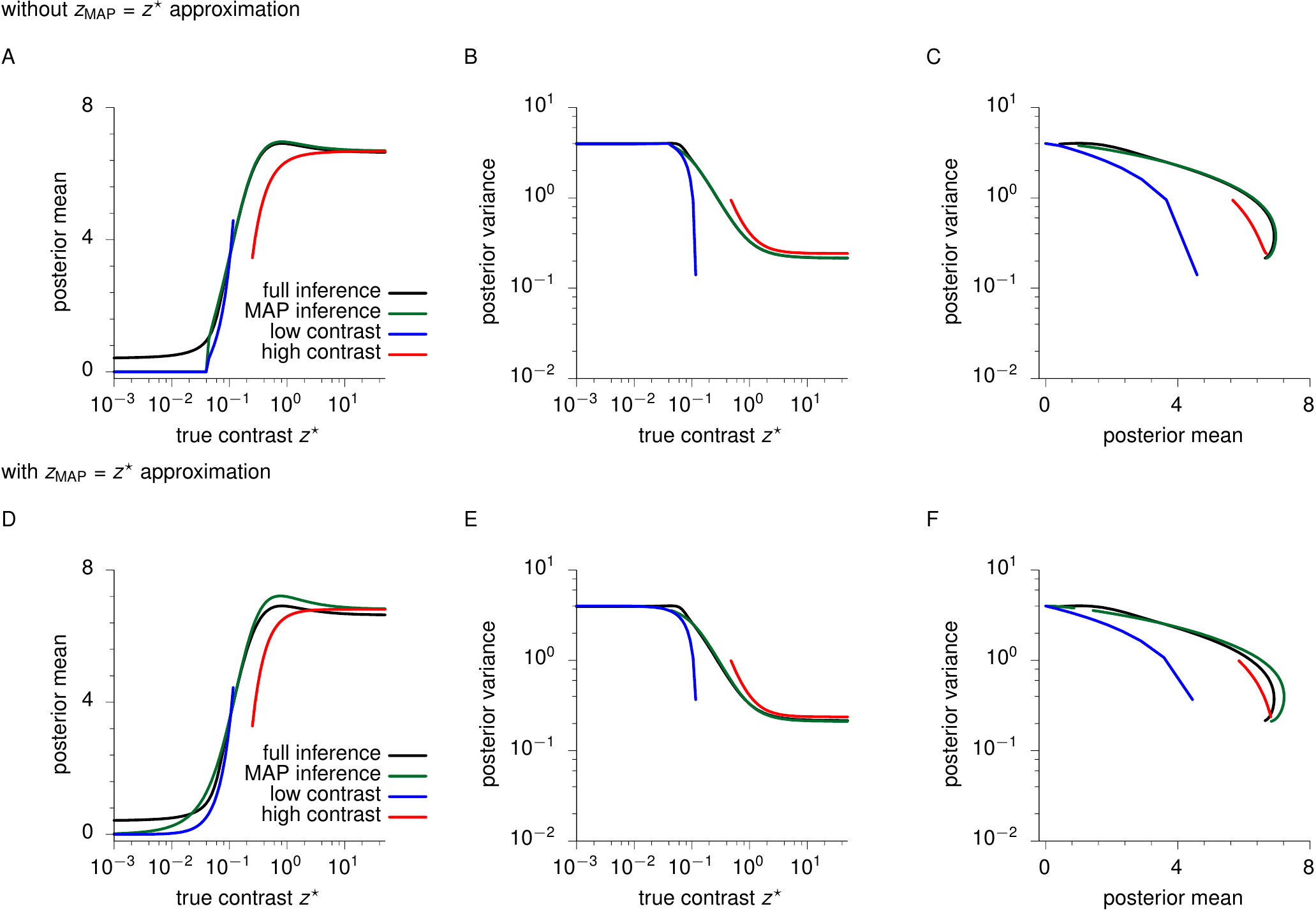}
\caption{\label{fig:undercomp}
Same as \Cref{fig:overcomp}, for the undercomplete case.}
\end{figure}

In order to explore the contrast-dependence of the mean and variance of the $\yy$-posterior, we included a single element in $\yys$ so that the corresponding mean and variance were scalars. Overall, we found that our approximations for both the low- and high-contrast limits were in good agreement with the full inference. In particular, they captured the qualitative dependence of both the mean and variance on contrast, as well as the way the mean and variance co-varied across different contrast levels (see \Cref{fig:overcomp,fig:undercomp} for the over-  and undercomplete cases, respectively).

\begin{figure}[t]
\centering
\includegraphics[height=0.2\textheight]{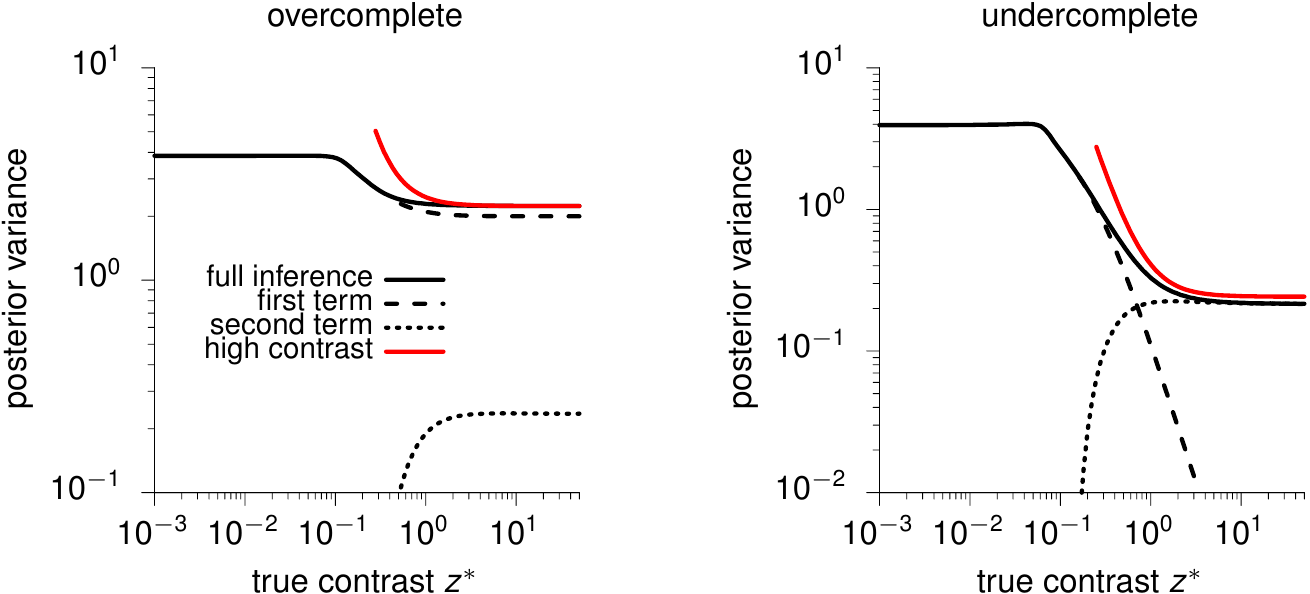}
\caption{\label{fig:variance}
Total posterior variance (black solid line), and the two terms contributing to it (\Cref{eq:postvar_full_inf}), the mean of the variance (black dashed line), and the variance of the mean (black dotted line), together with the high contrast approximation (\Cref{eq:postvars_HC}, with the $\zmap=\zs$ approximation, results were near-identical without this approximation), in the overcomplete (left) and undercomplete case (right).}
\end{figure}

We distinguished between two levels of approximation (see \Cref{sec:inferz}): one in which we took $\PP{\yy \given \xx} \simeq \PP{\yy \given \xx, z=\zmap}$ with $\zmap$ found via numerical optimization (\Cref{fig:overcomp,fig:undercomp}, panels A-C), and another one in which we assumed $\zmap\simeq\zs$, leading to $\PP{\yy \given \xx} \simeq \PP{\yy \given \xx, z=\zs}$ (\Cref{fig:overcomp,fig:undercomp}, panels D-F). Although, in general, posterior variability in $z$ was ignored by these approximations, for computing the posterior variance of $\yy$ in the high contrast regime, we applied a correction which did take it into account, and the consequent variance of the mean of $\yy$ (\Cref{sec:inferz,sec:lowhighcontrast}, \Cref{fig:variance}).

The second approximation had a more severe effect on the mean than on the variance, as the posterior variance under the GSM is independent of the true contrast, while $\zs$ enters the equations of the mean via $\xx$ (compare \Cref{eq:postmean,eq:postvar}). Thus, the asymptotic behavior of the mean in the second approximation was consistently off from that of the first approximation by the `correction' factor $\zs / \zmap$ (\Cref{fig:z_post}, bottom; cf.\ \Cref{eq:mean_LC_bef_approx,eq:mean_LC_aft_approx}, \Cref{eq:mean_HCO_bef_approx,eq:mean_HCO_aft_approx} and \Cref{eq:mean_HCU_bef_approx,eq:mean_HCU_aft_approx} in \Cref{sec:methods}). 
In particular, at low true contrasts, $\xx$ was dominated by the observation noise, so the $z$-posterior was dominated by the prior which had a peak at $0$, resulting in $\zmap=0$ for a finite range of true contrasts (\Cref{fig:z_post}, top). Consequently, the correction factor $\zs/\zmap$ diverged in the limit of small $\zs$ (\Cref{fig:z_post}, bottom). 
However, the assumption that the $z$-posterior is concentrated around $\zmap$ also broke down at low contrasts as it had considerable probability mass beyond $\zmap$ (\Cref{fig:z_post}, top) and so the full $\yy$-posterior behaved as if it was conditioned on a higher effective value of $z$ than $\zmap$ (e.g.\ its mean did not converge to $0$, and its variance did not converge to the prior variance as our analysis would have predicted). This meant that the second, seemingly more severe approximation, conditioning on $\zs$, which was consistently greater than $\zmap$ in this regime (\Cref{fig:z_post}, bottom), could in fact work better than the first one, conditioning on $\zmap$ (though it could still not predict the slightly above-zero mean at zero contrast).
More generally, we found that neither approximation introduced significant errors by itself, and that they had a particularly negligible effect in the range $0.1\leq \zs \leq 1.0$ over which the posterior mean and variance undergo most of their changes (\Cref{fig:overcomp,fig:undercomp}, black vs.\ green).

 
\clearpage

\section{Discussion}

By using simple approximations, we were able to study analytically the dependence of the posterior mean and variance in the GSM in the limit of low and high contrast. In both limits, we found they converged quadratically with contrast to their respective limiting values. Our numerical results show that the approximations are valid within a reasonable range, indeed providing practical validity. 
 
The characterization of the scaling of the mean and variance predicted by the GSM is highly relevant if it is to be applied to modeling neural data. While bottom up descriptions of neural dynamics, such as that provided by stabilized supralinear networks \citep{hennequin2016stabilized}, predict a dependence of the statistical moments of neural activity on contrast similar to that of the GSM posterior, the precise scaling each model predicts may not be identical. We have shown here how the GSM is a suitable candidate to model neural data in which both mean and variance saturate in the limits of low and high contrasts, and they do so in an approximately quadratic way.

\section{Methods}\label{sec:methods}
In the following we present the derivations for the scaling of the mean and variance in the limits of low and high contrast. In \Cref{eq:postvar_full_inf}, there are two contributions to the variance (due to the law of total variance), one coming from the expected value of the posterior variance given across contrast levels (which we will denote $\ZZone$), and second one resulting from the covariance of the mean produced by the variance in $z$ (which we will call $\ZZtwo$), so that:
\begin{align}
\ZZ &= \ZZone + \ZZtwo \label{eq:post_var_two_terms}
\intertext{where}
\ZZone &= \E{\ZZz}\\
&\simeq \ZZ^{\zmap}\\
&\simeq \ZZ^{\zs}\\
\intertext{assuming $\PP{z\given\xx}$ to be very narrow around $\zmap$, and}
\ZZtwo &= \Cov{\mm^z} \\
&\simeq \sz{\zmap} \, \frac{d\mmz}{dz}\biggr\rvert_{z=\zmap} \, \frac{d\mmz}{dz} \biggr\rvert_{z=\zmap}^\mathsf{T}  \label{eq:second_term_var_approx}\\
&\simeq \sz{\zs} \, \frac{d\mmz}{dz}\biggr\rvert_{z=\zs} \, \frac{d\mmz}{dz} \biggr\rvert_{z=\zs}^\mathsf{T}  \label{eq:second_term_var_approx}\\
\intertext{assuming $\mmz$ to be linear in $z$ over the relevant range of $z$ (where it has considerable probability mass), and where $\sz{z'}$ is the variance of $\PP{z\given\xx}$ which we further approximate by the `local variance' (i.e.\ the curvature of the logarithm) of $\PP{z\given\xx}$ around $z'$ (as in the Laplace approximation):}
1/\sz{z'} &\simeq -\frac{d^2 \ln \PP{z\given\xx}}{dz^2}\biggr\rvert_{z=z'}
\end{align}
As $\sz{\zmap}$, and especially $\sz{\zs}$, is finite and small in the low contrast limit but diverges in the high contrast limit (\Cref{fig:prior_like_post}), we only take $\ZZtwo$ into account in the latter case. Indeed, $\ZZone$ alone provides a very good approximation of $\ZZ$ in the low contrast regime (\Cref{fig:variance}).



\subsection*{Low contrast limit, \texorpdfstring{$z \rightarrow 0$}{z to 0}}

\begin{align}
\lim_{z \rightarrow 0} \ZZs^{\RN{1}}
&= \left[\CCs^{-1} + \frac{z^2}{\sx} \, \transpose{\AAsp} \left( \frac{z^2}{\sx} \, \RRsp + \II \right)^{-1} \, \AAsp \right]^{-1} \\
&\simeq \left[\CCs^{-1} + \frac{z^2}{\sx} \, \transpose{\AAsp} \left( \II - \frac{z^2}{\sx} \, \RRsp \right) \, \AAsp \right]^{-1} \\
&\simeq \left[\CCs^{-1} + \frac{z^2}{\sx} \, \transpose{\AAsp} \, \AAsp \right]^{-1} \\
&\simeq \CCs - \frac{z^2}{\sx} \, \CCs \, \transpose{\AAsp} \, \AAsp \, \CCs \\
&=\ZZszero - \frac{z^2}{\sx}\VV^{\LC} \label{eq:var_LC_bef_approx}\\
&\simeq\ZZszero - \frac{\zss}{\sx}\VV^{\LC}\label{eq:var_LC_aft_approx}\\
\text{with } \nonumber\\
\ZZszero &= \CCs\\
\VV^{\LC} &= \CCs \, \transpose{\AAsp} \, \AAsp \, \CCs\\ 
\intertext{~}
\lim_{z \rightarrow 0} \mms 
&= \frac{z}{\sx} \, \ZZs \, \transpose{\AAsp} \, \left( \frac{z^2}{\sx} \, \RRsp + \II \right)^{-1} \, \xx \\
&\simeq  \frac{z}{\sx} \, \ZZs \, \transpose{\AAsp} \, \left( \II - \frac{z^2}{\sx} \, \RRsp \right) \, \xx \\
&\simeq \frac{z}{\sx} \, \ZZs \, \transpose{\AAsp} \,  \xx \\
&\simeq \frac{z}{\sx} \, \CCs \, \transpose{\AAsp} \, \xx \\
&=  \frac{z\, \zs}{\sx} \, \CCs \, \transpose{\AAsp} \, \xxb \\
&=  \frac{\zs}{z} \left( \mmszero + \frac{z^2}{\sx}\MM^{\LC}\, \xxb \right) \label{eq:mean_LC_bef_approx}\\
&\simeq \mmszero + \frac{\zss}{\sx}\MM^{\LC}\, \xxb \label{eq:mean_LC_aft_approx}\\
\text{with } \nonumber \\
\mmszero &= \bf{0} \\
\MM^{\LC} &= \CCs \, \transpose{\AAsp}
\end{align}

We observe that for low contrasts $d\mm^z/dz \to 0$, while the variance of $z$ tends to the variance of the prior (finite), and therefore:

\begin{align*}
\lim_{z \rightarrow 0} \ZZs = \lim_{z \rightarrow 0} \ZZs^{\RN{1}}
\end{align*}

\subsection*{High contrast limit, \texorpdfstring{$z \rightarrow \infty$}{z to infinity}}

\subsubsection*{Overcomplete system ($\RRsp$ invertible)}

\begin{align}
\lim_{z \rightarrow \infty} \ZZs^{\RN{1}} 
&= \left[\CCs^{-1} + \frac{z^2}{\sx} \, \transpose{\AAsp} \left( \frac{z^2}{\sx} \, \RRsp + \II \right)^{-1} \, \AAsp \right]^{-1} \\
&= \left[\CCs^{-1} + \transpose{\AAsp} \left(\RRsp + \frac{\sx}{z^2} \, \II \right)^{-1} \, \AAsp \right]^{-1} \\
&\simeq \left[\CCs^{-1} + \transpose{\AAsp} \left(\RRsp^{-1} - \frac{\sx}{z^2} \, \RRsp^{-1}  \, \RRsp^{-1} \right) \, \AAsp \right]^{-1} \\
&= \left[\CCs^{-1} + \transpose{\AAsp} \, \RRsp^{-1}  \, \AAsp  - \frac{\sx}{z^2} \, \transpose{\AAsp} \, \RRsp^{-1}  \, \RRsp^{-1} \, \AAsp \right]^{-1} \\
&\simeq  \left(\CCs^{-1} + \transpose{\AAsp} \, \RRsp^{-1}  \, \AAsp\right)^{-1} + \frac{\sx}{z^2} \, \ZZsinf \, \transpose{\AAsp} \, \RRsp^{-1}  \, \RRsp^{-1} \, \AAsp \, \ZZsinf \\
&= \ZZsinf + \frac{\sx}{z^2}\VV^{\HC} \label{eq:var_HCO_bef_approx} \\
&\simeq {\ZZsinf}^{\RN{1}} + \frac{\sx}{\zss}{\VV^{\HC}}^{\RN{1}} \label{eq:var_HCO_aft_approx}\\
\text{with } \nonumber\\
{\ZZsinf}^{\RN{1}} &= \left(\CCs^{-1} + \transpose{\AAsp} \, \RRsp^{-1}  \, \AAsp\right)^{-1}\\
{\VV^{\HC}}^{\RN{1}} &= \ZZsinf \, \transpose{\AAsp} \, \RRsp^{-1}  \, \RRsp^{-1} \, \AAsp \, \ZZsinf\\
\intertext{~}
\lim_{z \rightarrow \infty} \mms 
&= \frac{z}{\sx} \, \ZZs^{\RN{1}} \, \transpose{\AAsp} \, \left( \frac{z^2}{\sx} \, \RRsp + \II \right)^{-1} \, \xx \\
&= \frac{\sx}{z^2} \, \frac{z}{\sx} \, \ZZs^{\RN{1}} \, \transpose{\AAsp} \, \left( \RRsp + \frac{\sx}{z^2} \, \II \right)^{-1} \, \xx \\
&\simeq \frac{1}{z} \, \ZZs^{\RN{1}} \, \transpose{\AAsp} \, \left( \RRsp^{-1} - \frac{\sx}{z^2} \, \RRsp^{-1} \, \RRsp^{-1} \right) \, \xx \\
&\simeq \frac{1}{z} \, \left( {\ZZsinf}^{\RN{1}} + \frac{\sx}{z^2} \, {\ZZsinf}^{\RN{1}} \, \transpose{\AAsp} \, \RRsp^{-1}  \, \RRsp^{-1} \, \AAsp \, {\ZZsinf}^{\RN{1}} \right) \, \transpose{\AAsp} \, \left( \RRsp^{-1} - \frac{\sx}{z^2} \, \RRsp^{-1} \, \RRsp^{-1} \right) \, \zs \, \xxb \\
&= \frac{\zs}{z} \left( {\ZZsinf}^{\RN{1}} + \frac{\sx}{z^2} \, {\ZZsinf}^{\RN{1}} \, \transpose{\AAsp} \, \RRsp^{-1}  \, \RRsp^{-1} \, \AAsp \, {\ZZsinf}^{\RN{1}} \right) \, \transpose{\AAsp} \, \left( \RRsp^{-1} - \frac{\sx}{z^2} \, \RRsp^{-1} \, \RRsp^{-1} \right) \, \xxb \\
&\simeq {\ZZsinf}^{\RN{1}} \, \transpose{\AAsp} \, \RRsp^{-1} \, \frac{\zs}{z} \, \xxb - \frac{\sx}{z^2} \, \left( {\ZZsinf}^{\RN{1}} \, \transpose{\AAsp} \, \RRsp^{-1} \, \RRsp^{-1} - {\ZZsinf}^{\RN{1}} \, \transpose{\AAsp} \, \RRsp^{-1}  \, \RRsp^{-1} \, \AAsp \, {\ZZsinf}^{\RN{1}} \, \transpose{\AAsp} \, \RRsp^{-1} \right) \, \frac{\zs}{z} \, \xxb \\
&= {\ZZsinf}^{\RN{1}} \, \transpose{\AAsp} \, \RRsp^{-1} \, \frac{\zs}{z} \, \xxb - \frac{\sx}{z^2} \, {\ZZsinf}^{\RN{1}} \, \transpose{\AAsp} \, \RRsp^{-1} \, \RRsp^{-1} \left( \II - \AAsp \, {\ZZsinf}^{\RN{1}} \, \transpose{\AAsp} \, \RRsp^{-1} \right)\, \frac{\zs}{z} \, \xxb \\
& = \frac{\zs}{z} \left(\mmsinf - \frac{\sx}{z^2}\MM^{\HC}\, \xxb \right)\label{eq:mean_HCO_bef_approx}\\
&\simeq \mmsinf - \frac{\sx}{\zss}\MM^{\HC}\, \xxb \label{eq:mean_HCO_aft_approx}\\
\text{with } \nonumber\\
\mmsinf& = {\ZZsinf}^{\RN{1}} \, \transpose{\AAsp} \, \RRsp^{-1} \, \xxb\\
\MM^{\HC} &= {\ZZsinf}^{\RN{1}} \, \transpose{\AAsp} \, \RRsp^{-1} \, \RRsp^{-1} \left( \II - \AAsp \, {\ZZsinf}^{\RN{1}} \, \transpose{\AAsp} \, \RRsp^{-1} \right)
\end{align}

\subsubsection*{Undercomplete system (\texorpdfstring{$\RRsp$}{Ro} non-invertible)}

Here, we will make use of \Cref{eq:Rchol} and note that if $\RRsp= \AAns \, \LL \, \LLT \, \AATns$ is low-rank and thus non-invertible it is still possible for $\LLT \, \AATns \, \AAns \, \LL$ to be full-rank and invertible (what we here call the undercomplete case), and so $\QQ$ is non-degenerate even in the $z \rightarrow \infty$ limit (see \Cref{sec:invert} for computing $\QQ = \left( \frac{z^2}{\sx} \, \RRsp + \II \right)^{-1}$ and its asymptotic form in this case).

\begin{align}
\lim_{z \rightarrow \infty} \ZZs^{\RN{1}} 
&= \left[ \CCs^{-1} + \frac{z^2}{\sx} \, \transpose{\AAsp} \left( \frac{z^2}{\sx} \, \RRsp + \II \right)^{-1} \, \AAsp \right]^{-1} \\
&= \left[ \CCs^{-1} + \frac{z^2}{\sx} \, \transpose{\AAsp} \, \QQ \, \AAsp \right]^{-1} \\
&\simeq \left[\CCs^{-1} + \frac{z^2}{\sx} \, \transpose{\AAsp} \, \left( \QQinf + \frac{\sx}{z^2} \, \QQb \right) \, \AAsp \right]^{-1} \\
&= \left[ \CCs^{-1} + \transpose{\AAsp} \, \QQb \, \AAsp + \frac{z^2}{\sx} \, \transpose{\AAsp} \, \QQinf \, \AAsp \right]^{-1} \label{eq:postcov_zinf_undercomp_interim} \\
&\simeq \frac{\sx}{z^2} \, {\VV^{\HC}}^{\RN{1}} \label{eq:var_HCU_bef_approx}\\
&\simeq \frac{\sx}{\zss} \, {\VV^{\HC}}^{\RN{1}}\\
&= {\ZZsinf}^{\RN{1}} + \frac{\sx}{\zss}{\VV^{\HC}}^{\RN{1}} \label{eq:var_HCU_aft_approx} \\
\text{with } \nonumber \\
{\ZZsinf}^{\RN{1}} &= \bf{0} \\
{\VV^{\HC}}^{\RN{1}} &= \left( \transpose{\AAsp} \, \QQinf \, \AAsp \right)^{-1}\\
\QQinf &= \II - \AAns \, \LL \, \left( \LLT \, \AATns \, \AAns \, \LL \right)^{-1} \, \LLT \, \AATns \\
\QQb &= \AAns \, \LL \, \left( \LLT \, \AATns \, \AAns \, \LL \right)^{-2}  \, \LLT \, \AATns  \\
\intertext{
As we shall see below, for deriving the asymptotic behaviour of $\mms$ we will need $\ZZs^{\RN{1}}$ up to higher order terms (as the $1/z^2$ term of $\ZZs^{\RN{1}}$ will cancel), for which we need a bit of extra work restarting from \Cref{eq:postcov_zinf_undercomp_interim}:}
\lim_{z \rightarrow \infty} \ZZs^{\RN{1}} &\simeq \frac{\sx}{z^2} \, \left[ \frac{\sx}{z^2} \, \left(\CCs^{-1} + \transpose{\AAsp} \, \QQb \, \AAsp \right) + \left({\VV^{\HC}}^{\RN{1}}\right)^{-1} \right]^{-1} \\
&\simeq \frac{\sx}{z^2} \, \left[ {\VV^{\HC}}^{\RN{1}} - \frac{\sx}{z^2} \, {\VV^{\HC}}^{\RN{1}} \, \left(\CCs^{-1} +  \transpose{\AAsp} \, \QQb \, \AAsp \right) {\VV^{\HC}}^{\RN{1}} \right] 
\end{align}

Now we are in the position to look at the asymptotic behaviour of the posterior mean.
\begin{align}
\lim_{z \rightarrow \infty} \mms 
&= \frac{z}{\sx} \, \ZZs^{\RN{1}} \, \transpose{\AAsp} \, \left( \frac{z^2}{\sx} \, \RRsp + \II \right)^{-1} \, \xx \\
&= \frac{z \, \zs}{\sx} \, \ZZs^{\RN{1}} \, \transpose{\AAsp} \, \QQ  \, \xxb \\
&\simeq \frac{z \, \zs}{\sx} \, \frac{\sx}{z^2} \, \left[ {\VV^{\HC}}^{\RN{1}} - \frac{\sx}{z^2} \, {\VV^{\HC}}^{\RN{1}} \, \left(\CCs^{-1} +  \transpose{\AAsp} \, \QQb \, \AAsp \right) {\VV^{\HC}}^{\RN{1}} \right] \, \transpose{\AAsp} \, \left( \QQinf + \frac{\sx}{z^2} \, \QQb \right)  \, \xxb \\
&= {\VV^{\HC}}^{\RN{1}} \transpose{\AAsp} \, \QQinf \, \frac{\zs}{z} \, \xxb - \frac{\sx}{z^2} \, {\VV^{\HC}}^{\RN{1}} \, \left[ \left(\CCs^{-1} +  \transpose{\AAsp} \, \QQb \, \AAsp \right) {\VV^{\HC}}^{\RN{1}} \, \transpose{\AAsp} \, \QQinf  - \transpose{\AAsp} \, \QQb \right] \, \frac{\zs}{z} \, \xxb \\
& = \frac{\zs}{z} \left(\mmsinf - \frac{\sx}{z^2}\MM^{\HC}\, \xxb\right) \label{eq:mean_HCU_bef_approx}\\
& \simeq \mmsinf - \frac{\sx}{\zss}\MM^{\HC}\, \xxb \label{eq:mean_HCU_aft_approx}\\
\text{where } \nonumber\\
\mmsinf &= {\VV^{\HC}}^{\RN{1}} \transpose{\AAsp} \, \QQinf \, \xxb\\
\MM^{\HC} &= {\VV^{\HC}}^{\RN{1}} \, \left[ \left(\CCs^{-1} +  \transpose{\AAsp} \, \QQb \, \AAsp \right) {\VV^{\HC}}^{\RN{1}} \, \transpose{\AAsp} \, \QQinf  - \transpose{\AAsp} \, \QQb \right]
\end{align}

\subsection*{Second term of the variance}

The derivation for the second term of the variance is lengthier and can be found in \Cref{sec:var_mean}, where we show that (both in the under- and overcomplete case):

\begin{align}
\ZZs^{\RN{2}} &\simeq \frac{\frac{\zss}{\zmap^2}}{ -n - \rank + 3 \rank \frac{\zss}{\zmap^2}} \left[ \mmsinf \mmsinfT +  3 \frac{\sx}{\zmap^2} \left(\mmsinf \xxbT \transpose{{\MM^{\HC}}}   + \MM^{\HC} \xxb\, \mmsinfT \right)\right]
\intertext{Where $\rank$ is the rank of $\AA \CC \AAT$, which, for $\AA$ matrices composed of orthogonal columns, like the ones we use here, will simply take the value $min(D_x,D_y)$.}
\intertext{For $\zmap \simeq \zs$}
\ZZs^{\RN{2}} &\simeq {\ZZsinf}^{\RN{2}} + \frac{\sx}{\zss}{\VV^{\HC}}^{\RN{2}}
\intertext{where}
{\ZZsinf}^{\RN{2}} &= \frac{1}{2 \rank- n} \mmsinf \mmsinfT\\
{\VV^{\HC}}^{\RN{2}} &= \frac{3}{2 \rank- n} \left(\mmsinf \xxbT \transpose{{\MM^{\HC}}}  + \MM^{\HC} \xxb\, \mmsinfT \right)
\end{align}
Where the second term of the variance shows the same type of scaling as the first term. We therefore can write: 
\begin{align}
\lim_{z \rightarrow \infty} \ZZs &\simeq \ZZsinf + \frac{\sx}{\zss}\VV^{\HC}
\intertext{where}
\ZZsinf &= {\ZZsinf}^{\RN{1}} + {\ZZsinf}^{\RN{2}}\\
\VV^{\HC}&= {\VV^{\HC}}^{\RN{1}} + {\VV^{\HC}}^{\RN{2}}
\end{align}
The final form of these expressions will depend on whether we work in the over- or undercomplete case, since the first term in each expression does.

\bibliographystyle{apalike}
\bibliography{scaling_mean_var_GSM}

\begin{thebibliography}{}

\bibitem[Coen-Cagli et~al., 2015]{coen2015flexible}
Coen-Cagli, R., Kohn, A., and Schwartz, O. (2015).
\newblock Flexible gating of contextual influences in natural vision.
\newblock {\em Nature neuroscience}.

\bibitem[Fiser et~al., 2010]{fiser10}
Fiser, J., Berkes, B., Orb\'an, G., and Lengyel, M. (2010).
\newblock Statistically optimal perception and learning: from behavior to
  neural representations.
\newblock {\em Trends Cogn Sci}, 14:119--30.

\bibitem[Hennequin et~al., 2016]{hennequin2016stabilized}
Hennequin, G., Ahmadian, Y., Rubin, D.~B., Lengyel, M., and Miller, K.~D.
  (2016).
\newblock Stabilized supralinear network dynamics account for stimulus-induced
  changes of noise variability in the cortex.
\newblock {\em bioRxiv}, page 094334.

\bibitem[Orb{\'a}n et~al., 2016]{orban2016neural}
Orb{\'a}n, G., Berkes, P., Fiser, J., and Lengyel, M. (2016).
\newblock Neural variability and sampling-based probabilistic representations
  in the visual cortex.
\newblock {\em Neuron}, 92(2):530--543.

\bibitem[Schwartz and Simoncelli, 2001]{schwartz2001natural}
Schwartz, O. and Simoncelli, E.~P. (2001).
\newblock Natural signal statistics and sensory gain control.
\newblock {\em Nature neuroscience}, 4(8):819--825.

\bibitem[Wainwright and Simoncelli, 1999]{wainwright1999scale}
Wainwright, M.~J. and Simoncelli, E.~P. (1999).
\newblock Scale mixtures of gaussians and the statistics of natural images.
\newblock In {\em Nips}, pages 855--861.

\end{thebibliography}

\clearpage
\appendix

\section*{Appendix}
\label{sec:Appendix}
\addcontentsline{toc}{section}{\nameref{sec:Appendix}}
\section{Deriving the low-dimensional posterior}\label{sec:lowpost}

The first step is to write the predictive distribution in terms of $\yys$ (marginalising out $\yyns$). This is easiest to do by rewriting \Cref{eq:pred} as
\begin{align}
\xx &= z \, \AA\, \yy + \sx \, \boldsymbol{\epsilon} &\boldsymbol{\epsilon} \sim \normal{\OO, \II} \\
&= z \, \AAs \, \yys + z \, \AAns \, \yyns+ \sx \, \boldsymbol{\epsilon}
\end{align}
importantly, here we treat $\yyns$ just as much as a random variable as $\boldsymbol{\epsilon}$, and its distribution  \emph{conditioned on $\yys$} has the following mean and covariance (knowing that the prior mean of both $\yys$ and $\yyns$ is $\OO$):
\begin{align}
\E{\yyns \given \yys} & = \CCsns \, \CCs^{-1} \, \yys \\
\cov{\yyns \given \yys} &= \CCns - \CCsns \, \CCs^{-1} \, \transpose{\CCsns} 
\intertext{As all our component distributions are normal, from this it follows that}
\xx \given \yys, z &\sim 
\normal{z \, \AAsp \, \yys, z^2 \, \RRsp + \sx \, \II} \\
\text{where } \nonumber \\
\AAsp &= \AAs + \AAns \, \CCsns \, \CCs^{-1} \\
\RRsp &= \AAns \, \left(\CCns - \CCsns \, \CCs^{-1} \, \transpose{\CCsns} \right) \, \AATns
\end{align}

Next, we rewrite the predictive distribution as an (unnormalised) distribution over $\yys$:
\begin{align}
\lefteqn{\normal{\xx; z \, \AAsp \, \yys, z^2 \, \RRsp + \sx \, \II} } \nonumber \\
&= \normal{z \, \AAsp \, \yys; \xx, z^2 \, \RRsp + \sx \, \II}\\
&\propto \normal{\yys; \lmms, \lZZs}\\
\lmms &= \frac{z}{\sx} \, \lZZs \, \transpose{\AAsp} \, \left( \frac{z^2}{\sx} \, \RRsp + \II \right)^{-1} \, \xx \\
\lZZs^{-1} &= \frac{z^2}{\sx} \, \transpose{\AAsp} \left( \frac{z^2}{\sx} \, \RRsp + \II \right)^{-1} \, \AAsp 
\end{align}

This allows us to derive the low-dimensional posterior as (c.f.\ \Crefrange{eq:posteriors}{eq:postvars})
\begin{align}
\yys \given \xx, z &\sim \normal{\mms, \ZZs}\\
\text{where } \mms 
&= \frac{z}{\sx} \, \ZZs \, \transpose{\AAsp} \, \left( \frac{z^2}{\sx} \, \RRsp + \II \right)^{-1} \, \xx \\
\text{and } \ZZs 
&= \left[\CCs^{-1} + \frac{z^2}{\sx} \, \transpose{\AAsp} \left( \frac{z^2}{\sx} \, \RRsp + \II \right)^{-1} \, \AAsp \right]^{-1} 
\end{align}

\section{Computing the inverse of \texorpdfstring{$\left( \frac{z^2}{\sx} \, \RRsp + \II \right)$}{the covariance for x} and its asymptotic form in the undercomplete case}\label{sec:invert}

This is relevant for computing both the posterior mean (\Cref{eq:postmeans}) and covariance (\Cref{eq:postvars}). By making use of the Cholesky decomposition of $\RRsp$ in \Cref{eq:Rchol} and the Woodbury identity, we obtain:
\begin{align}
\QQ 
&= \left( \frac{z^2}{\sx} \, \RRsp + \II \right)^{-1} \\
&= \left( \frac{z^2}{\sx} \, \AAns \, \LL \, \LLT \, \AATns + \II \right)^{-1} \\
&= \II - \frac{z^2}{\sx} \, \AAns \, \LL \, \left( \frac{z^2}{\sx} \, \LLT \, \AATns \, \AAns \, \LL + \II \right)^{-1} \, \LLT \, \AATns \\
&= \II - \AAns \, \LL \, \left( \LLT \, \AATns \, \AAns \, \LL + \frac{\sx}{z^2} \, \II \right)^{-1} \, \LLT \, \AATns 
\end{align}

We will be particularly interested in the asymptotic form of $\QQ$, which can be written as:
\begin{align}
\lim_{z \rightarrow \infty} \QQ 
&\simeq \II - \AAns \, \LL \, \left[ \left( \LLT \, \AATns \, \AAns \, \LL \right)^{-1} - \frac{\sx}{z^2} \, \left( \LLT \, \AATns \, \AAns \, \LL \right)^{-2}  \right] \, \LLT \, \AATns \\
&= \QQinf + \frac{\sx}{z^2} \, \QQb \\
\text{with } \QQinf &= \II - \AAns \, \LL \, \left( \LLT \, \AATns \, \AAns \, \LL \right)^{-1} \, \LLT \, \AATns \\
\QQb &= \AAns \, \LL \, \left( \LLT \, \AATns \, \AAns \, \LL \right)^{-2}  \, \LLT \, \AATns
\end{align}

We note that when $\RRsp= \AAns \, \LL \, \LLT \, \AATns$ is low-rank and thus non-invertible, it is still possible for $\LLT \, \AATns \, \AAns \, \LL$ to be full-rank and invertible (what we have here denoted the undercomplete case), and so $\QQ$ is non-degenerate even in the $z \rightarrow \infty$ limit.

\section{Numerical evaluation of \texorpdfstring{$\zmap$}{z MAP}}\label{sec:map}

We have considered in the present work two approaches to find the numerical value of $\zmap$, without substantial differences between them.

The first possibility is to perform a grid-search over $\mathcal{P}\left(z \given \xx \right) \propto \mathcal{P}\left(\xx \given z \right) \mathcal{P}(z)$, since we already have these values as computed for the colormap of \Cref{fig:z_post}. The drawback is that one needs to ensure to have a fine enough mesh around the peak value (of which one does not know the location \emph{a priori}) to find a reasonable value of $\zmap$.

An alternative is then to find the value of $z$ for which the first derivative of the posterior (or, for practicality, the log-posterior) vanishes, that is:
\begin{align}
\frac{\partial}{\partial z} ln \, \mathcal{P}\left(z \given \xx \right) &=   \frac{\partial}{\partial z} ln \, \mathcal{P}(z) +
\frac{\partial}{\partial z} ln \, \mathcal{P}\left(\xx \given z \right)
\end{align}
We have:
\begin{align}
ln \, \mathcal{P}(z) &= - n \, ln(z+b) + ctt.\\
\frac{\partial}{\partial z} ln \, \mathcal{P}(z) &= - \frac{n}{z+b}\\
\text{and} \nonumber\\
ln \, \mathcal{P}\left(\xx \given z \right) &= -\frac{1}{2} ln \, \left\lvert \SS(z)\right\rvert -\frac{1}{2}  \xxT \SS^{-1}(z) \xx + ctt.\\
\frac{\partial}{\partial z} ln \, \mathcal{P}\left(\xx \given z \right) &= -z \; Tr \left[\SS^{-1}(z)\left(\II -\xx \xxT \SS^{-1}(z)\right) \AA \CC \AAT\right] \label{eq:loglike_z_firt_deriv}\\
\text{with} \nonumber\\
\SS (z) &= \sigma_x^2 \II +  z^2 \AA \CC \AAT
\end{align}
Therefore, we look for the solution to the following 1D problem:
\begin{align}
\frac{\partial}{\partial z} ln \, \mathcal{P}\left(z \given \xx \right) =  
 - \frac{n}{z+b} - z \, Tr \left[\SS^{-1}(z)\left(\II -\xx \xxT \SS^{-1}(z)\right) \AA \CC \AAT\right] &\overset{!}{=} 0
\end{align}
This expression can then be fed into any root finding routine, to obtain a candidate $\zmap$. Since the derivative is not necessarily a monotonic function, one finally needs to check that the root thus found is a local maximum and not a minimum and, if so, whether it is truly a global maximum, also comparing the posterior there with the posterior at the $z = 0$ boundary.

\section{Variance produced by the variability of the mean}\label{sec:var_mean}

We recall from \Cref{eq:second_term_var_approx} that :
\begin{align}
\ZZs^{\RN{2}} &\simeq \frac{d\mm}{dz}\biggr\rvert_{z=\zmap} \frac{d\mm}{dz} \biggr\rvert_{z=\zmap}^\mathsf{T} \sz{\zmap}
\end{align}
From \Cref{eq:postmean}, we have:
\begin{align}
\frac{d\mm^z}{dz} &= \frac{d}{dz} \frac{z}{\sx} \, \ZZ^z \, \AAT \, \xx \\
&= \frac{\mm^z}{z} + \frac{z}{\sx} \, \frac{d\ZZ^z}{dz}  \, \AAT \, \xx \\
&= \frac{\mm^z}{z} - \frac{z}{\sx} \, \ZZ^z \frac{d}{dz}\left({\ZZ^z}^{-1}\right) \ZZ^z  \, \AAT \, \xx \\
&= \frac{\mm^z}{z} - \frac{2z}{\sx} \frac{z}{\sx} \, \ZZ^z  \AAT \AA \ZZ^z  \, \AAT \, \xx \\
&= \frac{\mm^z}{z} - \frac{2z}{\sx} \, \ZZ^z \AAT \AA\, \mm^z \\
&= \left(\frac{1}{z} \II- \frac{2z}{\sx} \, \ZZ^z \AAT \AA \right) \mm^z \\
\end{align}
If we now perform a Laplace approximation, we have: 
\begin{align}
\frac{\partial^2}{\partial z^2} ln \, \mathcal{P}\left(z \given \xx \right)\biggr\rvert_{z=z'} &\simeq - 1/\sz{z'}
\intertext{And therefore:}
\sz{\zmap} &\simeq - \left(\frac{\partial^2}{\partial z^2} ln \, \mathcal{P}\left(z \given \xx \right)\right)^{-1}_{z=\zmap}
\end{align}
From \Cref{eq:loglike_z_firt_deriv}, we then compute:
\begin{align}
\frac{\partial^2}{\partial z^2} ln \, \mathcal{P}\left(z \given \xx \right) &=  
\frac{\partial^2}{\partial z^2} ln \, \mathcal{P}\left(z\right) -\; Tr(\TT) - z \, Tr\left(\frac{\partial \TT}{\partial z}\right)
\intertext{with}
\TT &= \SS^{-1}(z)\left(\II -\xx \xxT \SS^{-1}(z)\right) \AA \CC \AAT\\
\frac{\partial \TT}{\partial z} &= -2z \, \SS^{-1} \left[ \AA \CC \AAT \TT - \xx \xxT \left(\SS^{-1} \AA \CC \AAT\right)^2 \right]
\intertext{and}
\frac{\partial^2}{\partial z^2} ln \, \mathcal{P}\left(z\right)  &= \frac{n}{(z +b)^2}
\end{align}

We are interested in the scaling of the variance with contrast, in the high contrast limit (since we know it vanishes for small $z$). To do that we need to compute an expansion of:
\begin{equation}
\SS^{-1} = \left( \sigma_x^2 \II +  z^2 \AA \CC \AAT \right)^{-1}
\end{equation}
noting that $\AA \CC \AAT$, is not necessarily invertible (in fact we know it will not be invertible in the undercomplete case). To get around this issue and provide a general expression, both for when $\AA \CC \AAT$ is invertible and when it is not, we begin by performing an eigenvalue decomposition of $\AA \CC \AAT$:
\begin{equation}
\AA \CC \AAT = \VV \DD \VVT
\end{equation}
where $\DD$ is a diagonal matrix containing the eigenvalues of $\AA \CC \AAT$ in descending order, and $\VV$ is the matrix whose columns are spanned by their corresponding eigenvectors. If we denote by $\rank$ the rank of $\AA \CC \AAT$, only the first $\rank$ elements in the diagonal of $\DD$ will therefore be non-zero.

We now proceed to decompose pixel space into two components, the first one, which we will denote the $\rank$ component will be the one corresponding to the first $\rank$ eigendirections, and the second one (which we will denote by $\bar{\rank}$), will be its orthogonal complement. In this way, $\DR$ corresponds to the non-zero block of $\DD$, and $\VV$ can be split into 
$\VR$ and $\VRB$. A block-wise inverse of $\SS$, in the limit of large contrasts, yields:
\begin{align}
\SS^{-1} &\simeq \SS^{-1}_\infty + \frac{1}{z^2} \tilde{\SS}\label{eq:S_inv_exp}
\intertext{where}
\SS^{-1}_\infty &= \sigma_x^2 \VRB \VRBT \\
\tilde{\SS} &= \VR \DR^{-1} \VRT
\intertext{We note that, in the overcomplete case, where $\AA \CC \AAT$ is full rank, \Cref{eq:S_inv_exp}, reduces to:}
\SS^{-1} &\simeq \frac{1}{z^2} \left(\AA \CC \AAT \right)^{-1}
\end{align}
Using the approximation:
\begin{align}
\xx \xxT &\simeq \zss \AA \CC \AAT
\intertext{we can rewrite $\TT$ as:}
\TT &\simeq \left(\SS^{-1}_\infty + \frac{1}{z^2} \tilde{\SS}\right) \left[\II - \zss \AA \CC \AAT \left(\SS^{-1}_\infty + \frac{1}{z^2} \tilde{\SS}\right) \right] \AA \CC \AAT\\
&= \frac{1}{z^2} \left( \ITL - \frac{\zss}{z^2} \ITL^2 \right)
\intertext{with $\ITL = \tilde{\SS} \, \AA \CC \AAT$ and 
$\SS^{-1}_\infty \, \AA \CC \AAT = \mathbf{0}$, since $\SS^{-1}_\infty$ spans only eigendirections corresponding to $0$ eigenvalues of $\AA \CC \AAT$. This means that, even though it would seem from \Cref{eq:S_inv_exp}, that $\SS^{-1}_\infty$ should play a dominant role in $\TT$ for large contrasts, it actually plays no role whatsoever. As a side comment, we note that in the overcomplete case $\ITL = \II$.}
\frac{\partial \TT}{\partial z} &\simeq \left(\frac{-2}{z^3} \ITL + 4 \frac{\zss}{z^5} \ITL^2 \right)
\intertext{Therefore}
-\; Tr(\TT) - z \, Tr\left(\frac{\partial \TT}{\partial z}\right) &= Tr \left[\frac{1}{z^2} \ITL - 3 \frac{\zss}{z^4} \ITL^2 \right]\\
&= \frac{1}{z^2} Tr (\, \ITL\, ) - 3 \frac{\zss}{z^4} Tr (\, \ITL^2 )= \frac{\rank}{z^2} \left( 1 - 3 \frac{\zss}{z^2} \right)
\intertext{Where we have used $Tr (\, \ITL\, ) = Tr (\, \ITL^2 ) = \rank$. So, for the power-law prior we here employ, and again in the high contrast limit}
\frac{\partial^2}{\partial z^2} ln \, \mathcal{P}\left(z \given \xx \right) &\simeq \frac{n}{(z + b)^2} + \frac{\rank}{z^2} \left( 1 - 3 \frac{\zss}{z^2} \right)\\
&\simeq \frac{n}{z ^2} + \frac{\rank}{z^2} \left( 1 - 3 \frac{\zss}{z^2} \right) = \frac{1}{z ^2} \left( n + \rank - 3 \rank \frac{\zss}{z^2} \right)
\intertext{And finally}
\sz{\zmap} &\simeq \frac{\zmap ^2}{ -n - \rank + 3 \rank \frac{\zss}{\zmap^2}} \label{eq:var_z_map}
\intertext{Which, for $\zmap \simeq \zs$ reduces to:}
\sz{\zmap} &\simeq \sz{\zs} \simeq \frac{\zss}{2 \rank -n} \label{eq:var_z_true}
\end{align}
For $\AA$ matrices composed of orthogonal columns, like the ones we use here, the rank $\rank$ will simply take the value $min(D_x,D_y)$.

Let's now look at the scaling of $d\mms$ for high contrasts. We have
\begin{align}
\mms^{\HC} &\simeq \frac{\zs}{z} \left(\mmsinf - \frac{\sx}{z^2}\MM^{\HC}\, \xxb\right)\\
\frac{d\mms^{\HC}}{dz} &\simeq \left(- \frac{\zs \mmsinf}{z^2}  + 3 \frac{\zs \sx}{z^4}\MM^{\HC}\, \xxb\right)
\intertext{and therefore}
\frac{d\mms^{\HC}}{dz}\biggr\rvert_{z=\zmap} = &\simeq \left(- \frac{\zs \mmsinf}{\zmap^2}  + 3 \frac{\zs \sx}{\zmap^4}\MM^{\HC}\, \xxb\right)
\intertext{Which, for $\zmap \simeq \zs$ becomes:}
\frac{d\mms^{\HC}}{dz}\biggr\rvert_{z=\zmap} = &\simeq \left(- \frac{\mmsinf}{\zs}  + 3 \frac{\sx}{\zsc}\MM^{\HC}\, \xxb\right)
\end{align}
We then have
\begin{align}
\left(\frac{d\mms^{\HC}}{dz}\frac{d\mms^{\HC}}{dz}^\mathsf{T}\right)\biggr\rvert_{z=\zmap} &\simeq 
\frac{\zss}{\zmap^4} \mmsinf \mmsinfT +  3 \frac{\zss \sx}{\zmap^6} \left(\mmsinf \xxbT \transpose{{\MM^{\HC}}}  + \MM^{\HC} \xxb\, \mmsinfT \right) \label{eq:dmu_map}
\intertext{And once again, for $\zmap \simeq \zs$:}
\left(\frac{d\mms^{\HC}}{dz}\frac{d\mms^{\HC}}{dz}^\mathsf{T}\right)\biggr\rvert_{z=\zmap\simeq \zs} &\simeq 
\frac{1}{\zss} \mmsinf \mmsinfT +  3 \frac{\sx}{\zsf} \left(\mmsinf \xxbT \transpose{{\MM^{\HC}}}  + \MM^{\HC} \xxb\, \mmsinfT \right)\label{eq:dmu_true}
\end{align}
Combining \Cref{eq:var_z_map,eq:dmu_map}, we obtain:
\begin{align}
\ZZs^{\RN{2}} &\simeq \frac{\frac{\zss}{\zmap^2}}{ -n - \rank + 3 \rank \frac{\zss}{\zmap^2}} \left[ \mmsinf \mmsinfT +  3 \frac{\sx}{\zmap^2} \left(\mmsinf \xxbT \transpose{{\MM^{\HC}}}  + \MM^{\HC} \xxb\, \mmsinfT \right)\right]
\intertext{Or, for $\zmap \simeq \zs$}
\ZZs^{\RN{2}} &\simeq {\ZZsinf}^{\RN{2}} + \frac{\sx}{\zss}{\VV^{\HC}}^{\RN{2}}
\intertext{where}
{\ZZsinf}^{\RN{2}} &= \frac{1}{2 \rank- n} \mmsinf \mmsinfT\\
{\VV^{\HC}}^{\RN{2}} &= \frac{3}{2 \rank- n} \left(\mmsinf \xxbT \transpose{{\MM^{\HC}}}  + \MM^{\HC} \xxb\, \mmsinfT \right)
\end{align}

\section{On the invertibility of \texorpdfstring{$\RRsp$}{Ro} and its rank}\label{sec:rank}

We know that for any real matrix $\MM$:
\begin{align}
\funrm{rank}{\MM\MMT} = \funrm{rank}{\MMT\MM} = \funrm{rank}{\MM} = \funrm{rank}{\MMT}
\end{align}

In particular, if $\MM = \AAns \, \LL$, we see that: 
\begin{align}
\funrm{rank}{\RRsp} = \funrm{rank}{\AAns \, \LL \, \LLT \, \AATns} = 
\funrm{rank}{\LLT \, \AATns \, \AAns \, \LL} = \funrm{rank}{\AAns \, \LL}
\end{align}

So if $\funrm{rank}{\AAns \, \LL} = r < min(\Nx,\Nyns)$ where $\AAns \LL \in \mathbb{R}^{\Nx \times \Nyns}$, then both $\AAns \, \LL \, \LLT \, \AATns$ and $\LLT \, \AATns \, \AAns \, \LL$ will be low rank and therefore not invertible. If this is the case, we can use neither the overcomplete nor the undercomplete approximation here presented in the high contrast regime.

\end{document}